\documentclass[twocolumn,prl,preprintnumbers,amsmath,amssymb,nofootinbib,superscriptaddress]{revtex4}

\usepackage{graphicx}
\usepackage{amssymb}
\usepackage{psfrag}
\usepackage{placeins}
\usepackage{graphicx}
\usepackage{diagbox}
\newcommand{\Lcdm}{\ensuremath{\Lambda}CDM~}

\newcommand{\rhob}{\ensuremath{\bar{\rho}}}
\newcommand{\Pb}{\ensuremath{\bar{P}}}
\newcommand{\grad}{\ensuremath{\vec{\nabla}}}

\newcommand{\omegazero}{\ensuremath{\omega^{(0)}_g}}
\newcommand{\omegasix}{\ensuremath{\omega^{(6)}_g}}
\newcommand{\omegafive}{\ensuremath{\omega^{(5)}_g}}
\newcommand{\LSS}{\ensuremath{*}}
\newcommand{\varw}{\emph{var-w}}
\newcommand{\constw}{\emph{const-w}}

\DeclareMathOperator\erf{erf\!}

\usepackage{color}

\begin{document}

\title{The Dark Matter equation of state through cosmic history}

\author{Michael Kopp}
\email{kopp@fzu.cz}
\affiliation{CEICO, Institute of Physics of the Czech Academy of Sciences, Na Slovance 2, Praha 8 Czech Republic}
\affiliation{Department of Physics, University of Cyprus, 1, Panepistimiou Street, 2109, Aglantzia, Cyprus}
\author{Constantinos Skordis}
\email{skordis@fzu.cz}
\affiliation{CEICO, Institute of Physics of the Czech Academy of Sciences, Na Slovance 2, Praha 8 Czech Republic}
\affiliation{Department of Physics, University of Cyprus, 1, Panepistimiou Street, 2109, Aglantzia, Cyprus}
\author{Daniel B Thomas}
\email{daniel.thomas-2@manchester.ac.uk}
\affiliation{Jodrell Bank Centre for Astrophysics, School of Physics \& Astronomy, The University of Manchester, Manchester M13 9PL, UK }
\affiliation{Department of Physics, University of Cyprus, 1, Panepistimiou Street, 2109, Aglantzia, Cyprus}
\author{St\'ephane Ili\'c}
\email{ilic@fzu.cz}
\affiliation{CEICO, Institute of Physics of the Czech Academy of Sciences, Na Slovance 2, Praha 8
Czech Republic}
\date{\today}

\begin{abstract}
Cold Dark Matter (CDM) is a crucial constituent of the current concordance cosmological model. 
Having a vanishing equation of state (EoS), its energy density scales with the inverse cosmic volume and  is thus uniquely described by a single number, its present abundance. 
We test the inverse cosmic volume law for Dark Matter (DM) by allowing its EoS to vary independently in eight redshift bins in the range $z=10^5$ and $z=0$.
We use the latest measurements of the Cosmic Microwave Background radiation from the Planck satellite and supplement them with
 Baryon Acoustic Oscillation (BAO) data from the 6dF and SDSS-III BOSS surveys, and with the Hubble Space Telescope (HST) key project data.
We find no evidence for nonzero EoS in any of the eight redshift bins. With Planck data alone, the DM abundance is most strongly constrained around matter-radiation 
equality $\omega^{\rm eq}_g =   0.1193^{+0.0036}_{-0.0035}$ (95\% c.l.), whereas its present day value is more weakly 
constrained $\omegazero = 0.16^{+0.12}_{-0.10}$ (95\% c.l.). Adding BAO or HST data does not significantly change the $\omega^{\rm eq}_g$ constraint, while
 $\omegazero$ tightens to $0.160^{+0.069}_{-0.065}  $ (95\% c.l.) and $0.124^{+0.081}_{-0.067}$ (95\% c.l.)  respectively.
Our results constrain for the first time the level of ``coldness'' required of the DM across various cosmological epochs and show that the DM abundance is strictly positive at all times.
\end{abstract}

\maketitle

\paragraph{Introduction}
Cosmological observations indicate that there is insufficient baryonic matter in the Universe
for the correct description of physical processes, if gravitational laws are dictated by General Relativity. 
A natural explanation is that most of the matter fields interact negligibly with light,
and are thus called Dark Matter, but  can still be seen through their gravitational effect.

Dark Matter (DM) is generally thought to be a  stable particle (or particles) not part of the standard model, however
it has so far remained elusive \cite{CDMS2016,PandaXII2016,XENON1002016,ADMX2016,Lux2017,CRESST2017,BrubakerZhongGurevichEtal2017}.
Cosmologically, it is usually modeled as Cold Dark Matter (CDM), which is part of the successful \Lcdm model 
that is consistent with observations of the Cosmic Microwave Background (CMB) (e.g. \cite{PlanckCollaborationXIII2015}), 
cosmic shear surveys (e.g. \cite{DESdataRelease1}), measurements of the background expansion such as BAO probes \cite{AndersonAubourgBaileyEtal2014},
  supernovae distance measurements \cite{JonesScolnicRiessEtal2017} and the observed abundance of light elements~\cite{PeimbertLuridianaPeimbert2007}.

The CDM model is defined by a phase space distribution function satisfying the collisionless Boltzmann equation with 
initially vanishing velocity dispersion and curl. This leads to a background CDM density $\rhob_c (a) \propto  a^{-3}$ ($a$ being the scale factor of the Universe) and 
equation of state (EoS) $w=0$ while the linearized density and velocity perturbations satisfy the continuity and pressureless Euler equations.\footnote{Note that we use $w$ to indicate the EOS of DM and not the EOS of Dark Energy which we assume to be -1 as in $\Lambda$CDM.} 
The resulting model arises naturally in the Weakly Interacting Massive Particle (WIMP) paradigm: the candidate particles are effectively 
collisionless and typically have an EoS  $w \sim 10^{-24} a^{-2}$ \cite{GreenHofmannSchwarz2005,Armendariz-PiconNeelakanta2014}, thus 
well described by CDM. The QCD axion is another CDM candidate~\cite{VisinelliGondolo2014}.

Not all DM candidates fit into the CDM paradigm, for instance, warm DM \cite{DodelsonWidrow1994,Armendariz-PiconNeelakanta2014,PiattellaCasariniFabrisEtal2015}, 
ultra light axions \cite{HuBarkanaGruzinov2000,HlozekGrinMarshEtal2015}, collisionless massive neutrinos \cite{ShojiKomatsu2003,LesgourguesTram2011}, 
 self-interacting massive neutrinos \cite{CyrRacineSigurdson2014,OldengottRampfWong2015}, Chaplygin gas \cite{SandvikTegmarkZaldarriagaEtal2004} and
self-interacting DM \cite{SpergelSteinhardt2000}. In addition, DM may interact with other species such as
neutrinos \cite{SerraZalameaCooray2010,WilkinsonBoehmLesgourgues2014}, photons \cite{BoehmEtAl2002,WilkinsonLesgourguesBoehm2014},
dark radiation \cite{Cyr-RacineSigurdson2012,DiamantiGiusarmaMena2013,Buen-AbadMarques-TavaresSchmaltz2015,LesgourguesMarques-TavaresSchmaltz2015} 
and Dark Energy \cite{Amendola2000,PourtsidouSkordisCopeland2013,D'AmicoHamillKaloper2016}.

Rather than taking the CDM description for granted we consider it timely to examine whether 
the data itself supports any deviation from the CDM paradigm, and thus to further determine or constrain DM properties. 
For our purpose we use the Generalized Dark Matter (GDM) model, first proposed by W. Hu \cite{Hu1998a}. 
The phenomenology  of the GDM model has been thoroughly investigated in \cite{KoppSkordisThomas2016}, where a connection was found with more fundamental theories, including
those of a rich self-interacting dark sector.
In addition, the recent work on the Effective Field Theory of Large Scale Structure (EFTofLSS) \cite{BaumannNicolisSenatoreEtal2012} suggests that, 
even for an initially pressureless perfect fluid, the non-linearities that develop on small scales affect  
the cosmological background and large scale linear perturbations, creating an effective pressure and viscosity such as those found in GDM.

The GDM model has been used to constrain DM properties with either constant or 
specific time dependences of the parameters \cite{Muller2005, CalabreseMigliaccioPaganoEtal2009, KumarXu2012, XuChang2013,ThomasKoppSkordis2016, KunzNesserisSawicki2016}. 
Here, we allow the DM EoS to vary more freely in time than  all previous studies.

\paragraph{The model}
We consider a flat Universe with only scalar perturbations, see \cite{KoppSkordisThomas2016} for more details and notation.
The background density $\rhob_g$ and pressure $\Pb_g$ of the DM evolve according to the conservation law
\begin{align} \label{GDMconservation}
\dot{\rhob}_g  = - 3 H (1+w) \rhob_g\,, \qquad \Pb_g=w \rhob_g\,,
\end{align}
where $H =\frac{\dot a}{a}$ is the Hubble parameter and the overdot denotes derivatives with respect to cosmic time $t$.
The parametric function $w(t)$ is freely specifiable with the case $w=0$ corresponding to a CDM background ($\rhob_g = \rhob_c$). 
The GDM model has two further free functions, the speed of sound, $c_s^2$, and the (shear) viscosity, $c_{\rm vis}^2$. 
The EoS $w$ is uncorrelated with the two perturbative parameters $c_s^2$ and $c_{\rm vis}^2$, as shown in \cite{ThomasKoppSkordis2016}, 
thus in this work we set these to zero and denote this class of GDM models by $w$DM. Consequently, replacing  CDM  by $w$DM  in the $\Lambda$CDM model
 leads to $\Lambda$$w$DM.

With this choice, the perturbed $w$DM fluid equations for the density contrast $\delta_g$ and velocity perturbation $\theta_g$ are given by
\begin{align} \label{GDMperts}
\frac{\dot{\delta}_g}{1+w}  &= 3   H \left( \frac{w \delta_g}{1+w}  +3 a H  c_a^2  \theta_g\right) -  \left(  \frac{1}{2} \dot{h}  -  \frac{1}{a}\grad^2\theta_g \right) \notag 
\\
\dot{\theta}_g & = -  H  \theta_g \,,\qquad  c_a^2 = \frac{\dot{\Pb}_g}{\dot{\rhob}_g} = w - \frac{\dot{w}}{3 H (1+w)} \text{.}
\end{align}
Here, $c_a^2$ is the adiabatic speed of sound and $h$ is a metric perturbation in synchronous gauge \cite{Hu1998a,KoppSkordisThomas2016}.
The Euler equation $\dot{\theta}_g  = -  H  \theta_g$ is identical to that of CDM, which implies the solution $\theta_g=0$.
An example of $w$DM is the combination of CDM and $\Lambda$ interpreted as a single fluid with $w=-(1+\rhob_c/\rhob_\Lambda)^{-1}$. 
A large degeneracy between $\Omega_\Lambda$ and $w$ is thus expected at late times (see also \cite{TutusausEtal2016}).

\paragraph{Methodology}
The $w$DM fluid equations \eqref{GDMconservation}, \eqref{GDMperts} were implemented in the Boltzmann code CLASS \cite{BlasLesgourguesTram2011} 
as in~\cite{KoppSkordisThomas2016,ThomasKoppSkordis2016}.  A sufficiently general time-dependence of $w$ was achieved by binning its evolution into $N=8$ scale factor bins, 
whose edges are $\tilde a_i = 10^{\{0,-1,-1.5,-2,-2.5,-3,-3.5,-4\}}$. 
The bins were smoothly connected using $w(a) =  \frac{w_i-w_{i+1} }{2} \erf\left( \frac{\ln (a/\tilde a_{i+1})}{\sigma_a}\right) + \frac{ w_i+w_{i+1}}{2}$ 
for $a_{i+1}<a<a_i$, with bin centers $a_i=\sqrt{ \tilde a_{i} \tilde a_{i+1}}$ for $1\leq i\leq N-2$ while $a_0=1$ and $a_{N-1} = 0$.
Because of the aforementioned degeneracy of $w$DM with CDM and $\Lambda$, we chose a wider bin in the late Universe. 
 
The $\sigma_a$ parameter controls the transition width between bins; it was set to $1/20$ so that the transition is small compared to the bin width.
We tested that this choice does not affect our conclusions. 

We define a dimensionless scaled $w$DM density 
\begin{equation} \label{Defomegag}
\omega_g \equiv a^3 \rhob_g\,\frac{8 \pi G}{3\times (100\, \rm{km/s/Mpc})^2}\text{.}
\end{equation}
When $w=0$ through cosmic history, $\omega_g $ is a constant equal to the conventional dimensionless CDM density $\omega_c$.
In general however, $\omega_g$ varies over time and is fully determined by the $N+1$ parameters $\omegazero, w_i$. 
We use the notation  $\omega_g^{(i)}=\omega_g(a_i)$  and similarly for other functions with subscripts, so that the present day DM abundance 
is $\omegazero = \omega_g(a_0)$.  For functions without a subscript we instead write $H_i = H(a_i)$ and $w_i=w(a_i)$.

Our parameter constraints were obtained as in \cite{ThomasKoppSkordis2016} and we present only brief details here. We used the Markov chain Monte Carlo 
code MontePython~\cite{AudrenLesgourguesBenabedetal2013} and established convergence of the chains using the Gelman-Rubin criterion~\cite{GelmanRubin1992}.
Our total parameter set 
 \begin{equation}
 (\omega_b, \omegazero, H_0,n_s, \tau, \ln 10^{10} A_s, w_i)
 \end{equation}
consists of 6 $\Lambda$CDM parameters and the 8 values $w_i$. We denote the $\Lambda w$DM model with 8 bins as ``\varw'' and 
 the previously studied model~\cite{ThomasKoppSkordis2016} with $w=\,$const as ``\constw''.
We assumed adiabatic initial conditions.

\begin{figure}[t!]
\begin{center}
\includegraphics[width=0.48 \textwidth]{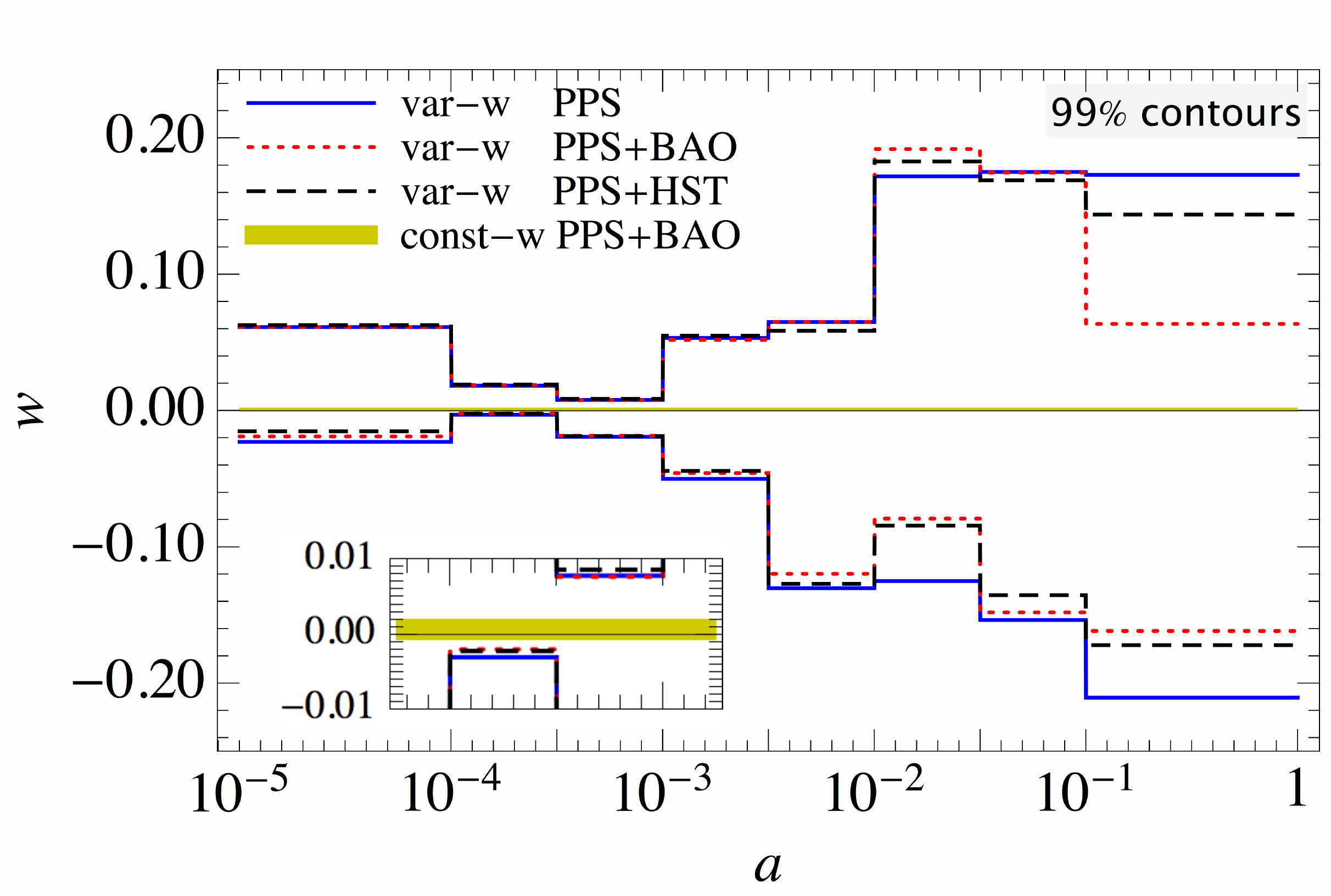}
\end{center}
\caption{99\% confidence regions on the EoS of DM, $w(a)$ for $\sigma_a=0$. 
The 8 bins are indicated by the large ticks on the $a$-axis.
The different line styles correspond to different data sets and models specified in the legend. 
The inset shows the region between $\tilde a_7=10^{-4}$ and $\tilde a_5=10^{-3}$ magnified.
Within the \constw~narrow stripe lies the \Lcdm model indicated by the black solid line.
}
\label{varwovertime_THREE}
\end{figure}

We used the Planck 2015 data release \cite{PlanckCollaborationXI2015} of the CMB anisotropies power spectra, composed of the low-$l$ T/E/B likelihood and the 
full TT/TE/EE high-$l$ likelihood with the complete ``not-lite'' set of nuisance parameters.
~\footnote{For full details, see the Planck papers and wiki http://wiki.cosmos.esa.int/planckpla2015/index.php/.} 
These likelihoods combined are referred to as Planck Power Spectra (PPS). 
We also added selectively the HST key project prior on $H_0$~\cite{RiessMacriCasertanoEtal2011},
BAO from the 6dF Galaxy Survey~\cite{BeutlerBlakeCollessEtAl2011} and the Baryon Oscillation Spectroscopic Survey Sloan Digital Sky Survey~\cite{AndersonAubourgBaileyEtal2014}, and the 
Planck CMB lensing likelihood (respectively referred to as HST, BAO and Lens thereafter).

We set uniform priors on $\tau$ and $H_0$ such that $0.01<\tau$ and $45 \leq H_0 \leq 90$ respectively. We used the same priors on Planck nuisance parameters
  and the same neutrino treatment as in \cite{ThomasKoppSkordis2016}. The helium fraction was set to $Y_{\rm He}=0.24667$ \cite{PlanckCollaborationXIII2015}.

\begin{figure}[t!]
\begin{center}
\includegraphics[width=0.48 \textwidth]{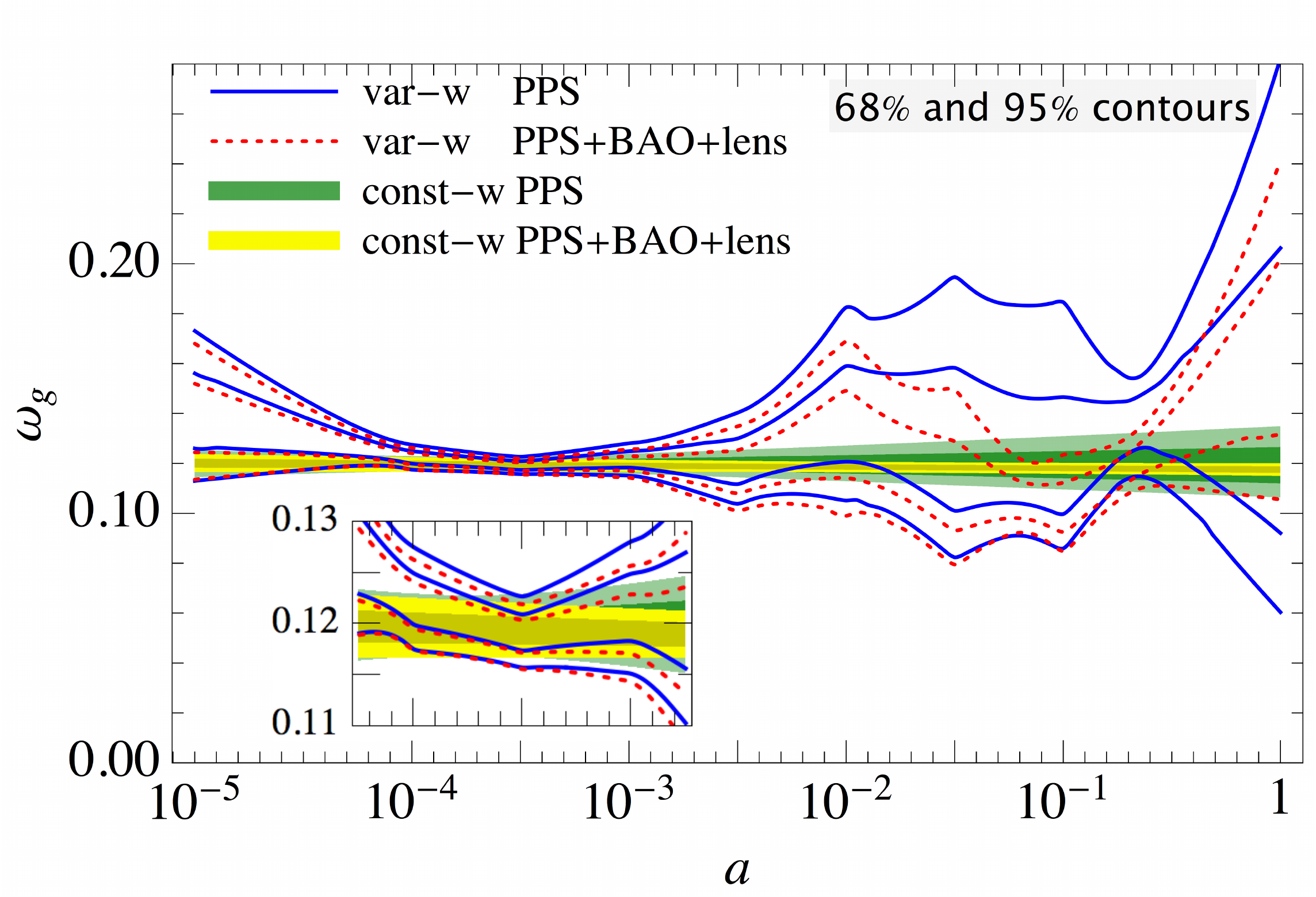}
\end{center}
\caption{
The $68\%$ and $95\%$ contours of the 1D marginalized posteriors on the DM abundance $\omega_g(a)$.
We show the \varw~and \constw~models, both with two different data sets (PPS and PPS+BAO+Lens) as specified in the legend.
}
\label{rhoovertime}
\end{figure}

\paragraph{Results}
Our main results are constraints on the time dependence of DM EoS $w(a)$ and  abundance $\omega_g(a)$ shown in Figs.\,\ref{varwovertime_THREE} and  \ref{rhoovertime}.
For comparison, we also show the constraints on the \constw~model already discussed in \cite{ThomasKoppSkordis2016}. 
We list the 95\% confidence regions of all parameters in Table \ref{table_results}.

In Fig.\,\ref{varwovertime_THREE} we observe that \Lcdm lies in the 99\% confidence region of the \constw~model, which in turn lies in the 99\% confidence region of the \varw~model, 
such that the constraints are nested like the models themselves.
There is no evidence for significant deviations of the DM EoS from 0 at any time. Consequently, any model selection criteria will favor \Lcdm\!\!.

 The constraints on $w$ are the strongest between $a_6$ and $a_5$ enclosing the matter-radiation equality $a_{\rm eq} \simeq 3\times10^{-4}$,
and are about a factor 2 weaker compared to the \constw~model.
In other bins the constraints on $w$ weaken significantly.   Adding 
the BAO or HST dataset has only a minor effect on \varw~constraints and only 
tightens limits in the rightmost bin. 
 As was the case for the \constw~model, \cite{ThomasKoppSkordis2016}, 
adding CMB lensing does not significantly  improve the constraints.

Let us now compare in more detail the DM abundance $\omega_g(a)$ of the \varw~and \constw~models focussing only on the two dataset combinations PPS and PPS+BAO+Lens.
In Fig.\ref{rhoovertime} we see that, like $w(a)$, $\omega_g$ is most tightly constrained between $a_6=10^{-3.75}$ and $a_5=10^{-3.25}$, in fact almost as tightly as for the \constw~model (see inset in
 Fig.\ref{rhoovertime}).  Around $a=0.4$ there is a squeeze in the constraints of $\omega_g$ from PPS, which extends to $a\sim (0.08,0.4)$ when BAO  or HST are included. 
 At all times a vanishing DM abundance ($\omega_g=0$) is inconsistent with the data.
More quantitatively, we find for the \varw~model at 95\% c.l. $\omega^{\rm eq}_g =   0.1193^{+0.0036}_{-0.0035}$ and $\omegazero =0.16^{+0.12}_{-0.10}$ 
with PPS only, whereas for PPS+BAO+Lens we get $\omega^{\rm eq}_g =   0.1189^{+0.0032}_{-0.0033}$  and $\omegazero = 0.169^{+0.067}_{-0.065}$. 
For \constw-PPS+BAO+Lens we find $\omega^{\rm eq}_g=0.1193^{+0.0026}_{-0.0026}$, whereas the \Lcdm result is $\omega_c=\omega^{\rm eq}_g=0.1184^{+0.0022}_{-0.0022}$, 
see \cite{ThomasKoppSkordis2016}.

\begin{figure}[t!]
\includegraphics[width=0.48 \textwidth]{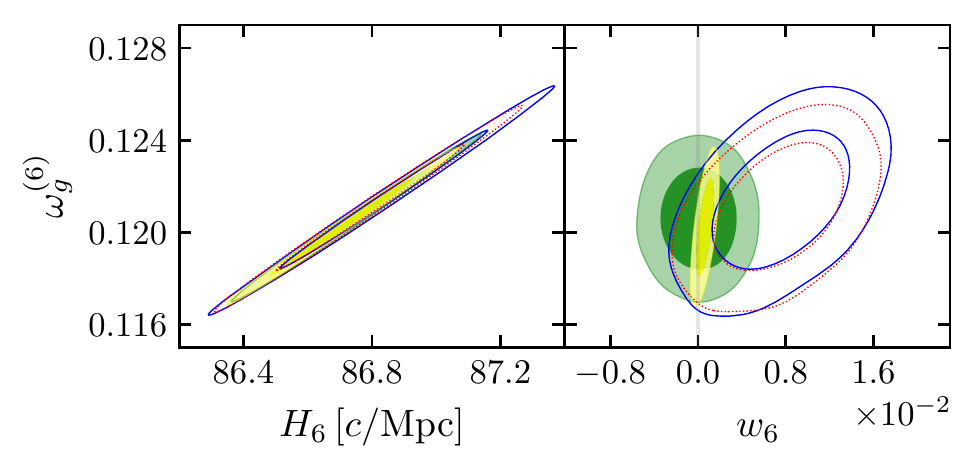}
\caption{68\% and 95\% contours of 2D marginalized posteriors of $\omegasix$-$H_6$ (left) and  $\omegasix$-$w_6$ (right). The line styles and colors 
are as in Fig.\ref{rhoovertime} and Fig.\ref{Contours_late}}
\label{Contours_early}
\end{figure}

Consider the tightly constrained region around $a_{\rm eq}$, $a_6<a<a_5$, as shown in Fig.\ref{rhoovertime}  (see also the inset in the same figure). 
As discussed in~\cite{Hu1998a,KoppSkordisThomas2016} the GDM abundance $\omega_g(a)$ and expansion rate $H(a)$ in the early Universe determine
the time of matter radiation equality and thereby the amount of potential decay until recombination. This in turn sets the relative 
heights of the first few peaks of the CMB temperature angular power spectrum. Both the \constw~and \varw~models constrain $\omega_g$ 
around  $a_{\rm eq}$ at a similar level (see above). The degeneracy between $H_0$ and $\omega_g$ in the \constw~model~\cite{ThomasKoppSkordis2016} 
translates into a degeneracy between $H_6$ and $\omegasix$ in the \varw~model as seen in the left panel of Fig. \ref{Contours_early}. 
Indeed, the $H_6$-$\omegasix$ contours reveal how well the CMB constrains a combination of the expansion rate and  the abundance of DM  around $a_{\rm eq}$.
The degeneracy between $w$-$\omega_g$~\cite{ThomasKoppSkordis2016,KoppSkordisThomas2016} in the \constw~model due to the same effect is also seen as
a degeneracy between  $w_6$ and $\omegasix$ (right panel of Fig. \ref{Contours_early}), however, in the \varw~model it is weakened as $w_6$ has only an indirect effect on
$a_{\rm eq}$, contrary to $\omegasix$.  Similar correlations exist for $H_5 -\omegafive$ and $\omegafive - w_5$ but in the opposite direction.

\begin{figure}[t!]
\includegraphics[width=0.5 \textwidth]{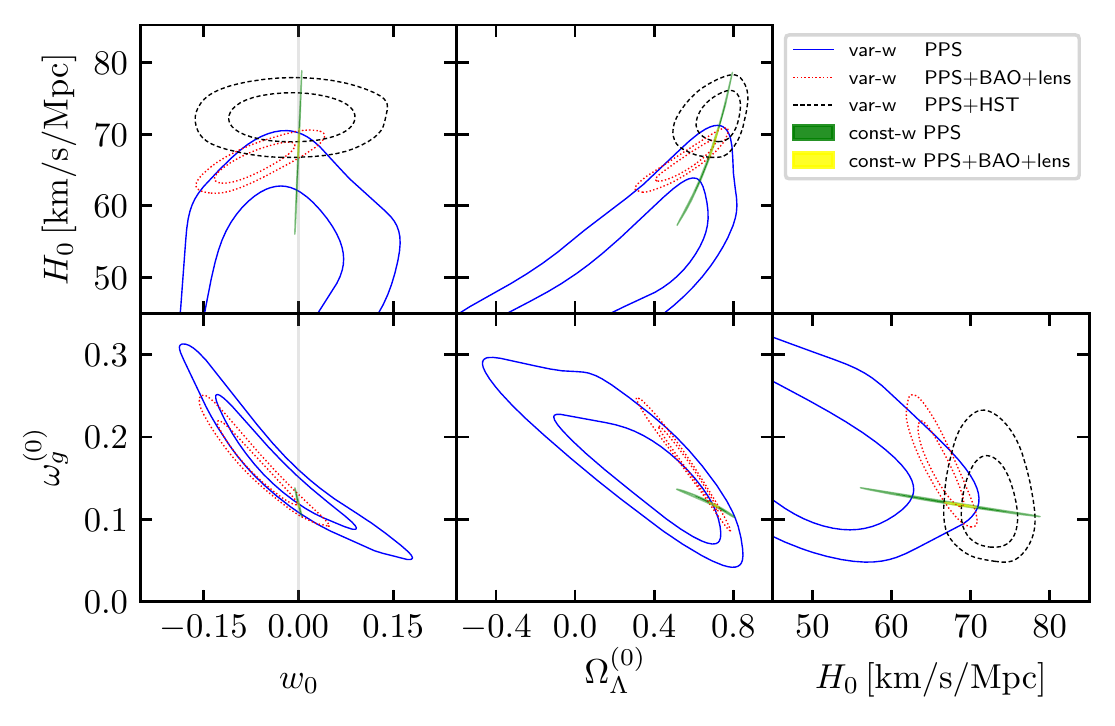}
\caption{68\% and 95\% contours of 2D marginalized posteriors for combinations of parameters in the set $\{H_0,w_0,\omegazero,\Omega_\Lambda\}$. The PPS+HST contours in the $\omegazero$-$w_0$ and $\omegazero$-$\Omega^{(0)}_\Lambda$ panels are not displayed as they are very similar to PPS+BAO(+lens).}
\label{Contours_late}
\end{figure}

The origin of the squeeze around $a\sim (0.08-0.4)$ is of an entirely different nature.
 The angular diameter distance $d^{\LSS}_A$ to the last scattering surface is given by 
 \begin{equation} \label{d_ang}
 d^{\LSS}_A  = a_{\LSS} \int_{a_{\LSS}}^1 d \ln a\, (aH)^{-1} = a_{\LSS} (\eta_0 - \eta_{\LSS}) 
\,.
 \end{equation}
Here, the second equality has been written in terms of the conformal time ($\eta = \int dt/a$) today, $\eta_0$, and at the last scattering surface, $\eta_\LSS$.
The largest contribution to  $d_A^\LSS$ comes from the first $\ln a$ bin where $\eta$ grows from $\sim 4000\,$Mpc to $\sim 14000\,$Mpc
and constitutes $\sim  70\%$ of the total.
 This contribution can be strongly constrained by geometric probes. 
Within the first scale factor bin, we have  $H =  \sqrt{\omegazero (a^{-3 (1+w_0)}-1)+H_0^2}$ (with $H$ and $H_0$ in units of $100\,\mathrm{km/s/Mpc}$).
Hence for a one-parameter family of $\omegazero$ and $w_0$ the combination  $(aH)^{-1}$ is
 approximately constant. As this is the largest contributor to $d^{\LSS}_A$ we expect $w_0$ and $\omega^{(0)}_{g}$ to be anticorrelated, as is indeed
observed in Fig.\ref{Contours_late} (lower left panel).
The inclusion of BAO or HST data significantly improves and shifts the constraints on $\eta(a)$ and is in turn reflected in the $\omega_g$ and $w$ constraints.

For the \varw~model, PPS alone allows for very low $H_0$, as low as $45\,$km/s/Mpc, corresponding to our hard prior on $H_0$, see the blue contours in the top left and bottom right panels of  
Fig.\ref{Contours_late}.  Adding BAO (red dotted lines) or HST (black dashed lines) shrinks the posteriors and also moves the mean of $H_0$ back towards the range consistent with the
 \constw~model \cite{ThomasKoppSkordis2016}. BAO (and also HST) data leads to a degeneracy between $H_0$ and $w_0$ and between $H_0$ and $\omegazero$, as in the \constw~model. However, 
as the present day values of $H$ and $\omega_g$ are no longer anchored to their early Universe values,
the degeneracy axis is rotated and the contours are not as flattened. 

In the middle panels of Fig.\ref{Contours_late} we display the 2D marginalized posteriors of the $H_0$-$\Omega_\Lambda$ and $\omegazero$-$\Omega_\Lambda$ planes.
As $\omega_b$  is well constrained, $w$DM and $\Lambda$ are the only relevant species in the late (flat) Universe and
are expected to have their abundances anticorrelated.  Indeed, the parameter $\omegazero$ is anticorrelated
with  $\Omega^{(0)}_\Lambda$ for all data sets. 
The combination of CDM and $\Lambda$ may be modeled by $w$DM; in that model, however,  $w$ changes steeply only within the $w_0$ bin so that this behavior is unaffected.
When BAO or HST data are included the slope and size of the contours change strongly as the late Universe behavior 
dissociates from the early Universe in the \varw~model. The negative values of $\Omega^{(0)}_\Lambda$ are correlated with the low values of $H_0$,
and, while allowed by PPS, they disappear when $H_0$ is better constrained after including BAO or HST data.

\begin{table}
{\tiny
\begin{tabular} {| l | c | c | c |}
\hline
\backslashbox[17mm]{\!Parameter}{Data} &  PPS & PPS+BAO & PPS+HST\\
\hline
\hline
{$100\omega_b    $} & $2.221^{+0.041}_{-0.040}   $ & $2.217^{+0.040}_{-0.038}   $     & $2.218^{+0.039}_{-0.038}   $\\
\hline
{$\omegazero       $} & $0.16^{+0.12}_{-0.10}      $  &  $0.160^{+0.069}_{-0.065}   $  & $0.124^{+0.081}_{-0.067}   $\\
\hline
{$H_0 [\mathrm{km/s/Mpc}]       $} & $< 65.9                    $   &  $66.6^{+3.7}_{-4.0}        $   & $72.3^{+4.5}_{-4.6}        $\\
\hline
{$n_s            $} & $0.974^{+0.021}_{-0.020}   $   &  $0.974^{+0.020}_{-0.020}   $  & $0.977^{+0.020}_{-0.020}   $\\
\hline
{$\tau$} & $0.072^{+0.039}_{-0.035}   $   & $0.076^{+0.040}_{-0.034}   $   & $0.076^{+0.037}_{-0.034}   $\\
\hline
{$\ln(10^{10} A_s)$} & $3.085^{+0.076}_{-0.069}   $  &  $3.096^{+0.078}_{-0.067}   $  & $3.096^{+0.073}_{-0.067}   $\\
\hline
{$w_0            $} & $-0.03^{+0.17}_{-0.14}     $    & $-0.056^{+0.091}_{-0.083}  $  & $-0.01^{+0.12}_{-0.13}     $\\
\hline
{$w_1            $} & $0.01^{+0.13}_{-0.13}      $    & $0.02^{+0.12}_{-0.13}      $ & $0.02^{+0.12}_{-0.12}      $\\
\hline
{$w_2            $} & $0.02^{+0.12}_{-0.11}      $    & $0.05^{+0.10}_{-0.10}      $  & $0.05^{+0.11}_{-0.10}      $\\
\hline
{$w_3            $} & $-0.044^{+0.075}_{-0.072}  $    & $-0.036^{+0.072}_{-0.068}  $  & $-0.041^{+0.075}_{-0.067}  $\\
\hline
{$w_4            $} & $0.002^{+0.038}_{-0.039}   $    & $0.005^{+0.036}_{-0.038}   $  & $0.005^{+0.038}_{-0.038}   $\\
\hline
{$w_5            $} & $-0.006^{+0.011}_{-0.010}  $    & $-0.006^{+0.010}_{-0.010}  $  & $-0.005^{+0.011}_{-0.010}  $\\
\hline
{$w_6            $} & $0.0078^{+0.0079}_{-0.0081}$    & $0.0084^{+0.0078}_{-0.0079}$  & $0.0085^{+0.0080}_{-0.0080}$\\
\hline
{$w_7            $} & $0.021^{+0.031}_{-0.032}   $    &  $0.022^{+0.030}_{-0.031}   $     & $0.025^{+0.029}_{-0.030}   $\\
\hline
\hline
{$\Omega_\Lambda^{(0)} $} & $0.34^{+0.45}_{-0.58}      $   & $0.58^{+0.18}_{-0.21}      $   & $0.72^{+0.14}_{-0.16}      $\\
\hline
{$\sigma_8       $} & $0.71^{+0.45}_{-0.36}      $    & $0.72^{+0.27}_{-0.23}      $  & $0.91^{+0.43}_{-0.39}      $\\
\hline
\end{tabular}
}
\caption{95\% confidence intervals of \varw~parameters.}
\label{table_results}
\end{table}

\paragraph{Implications}
In the $w$DM model the DM abundance $\omega_g$ may deviate from its expected (constant) CDM value throughout cosmic history,
causing only minimal changes to the clustering properties of DM. 
Hence, the constraints on $w$ and $\omega_g$ are conservative. 
One could also conservatively allow for general $c_s^2(a,k)$ and $c_{\rm vis}^2(a,k)$ and marginalize over them. 
However, as $w$ is almost uncorrelated 
 with $c_s^2$ and $c_{\rm vis}^2$, we expect such procedure to give constraints similar to those here. 
In the cases of warm DM and EFTofLSS the parameters $w$, $c_s^2$ and $c_{\rm vis}^2$ are interrelated so that the $w$ constraints will be driven
 by $c_s^2$ and $c_{\rm vis}^2$, and hence, tightened further~\cite{Armendariz-PiconNeelakanta2014, KunzNesserisSawicki2016}.
Adding spatial curvature and/or neutrino mass would likely widen the $\omega_g$ constraints on the squeeze at $a\sim\{0.08-0.4\}$~\cite{FerreiraSkordisZunckel2008} 
and in the latter case on the tightly constrained region around $a_{\rm eq}$ as well.

When applying our constraints to generic theories of Dark Matter, including those coming from modifications of gravity, one must keep in mind our underlying assumption of adiabaticity.
As models of modified gravity will typically have additional fields leading to more types of isocurvature modes, we expect our constraints to be less applicable in those cases.
However, within our adiabatic assumption we expect our constraints to be valid for any theory of Dark Matter or modified gravity. The cosmological background in any such theory 
will have to evolve as in $\Lambda$CDM (see for example \cite{BanadosFerreiraSkordis2009}), around matter radiation equality and before decoupling. 
Typical examples include DM-DE coupled models~\cite{Amendola2000,PourtsidouSkordisCopeland2013,D'AmicoHamillKaloper2016}. Explicit realizations where a CDM-like
 background decays into DE are given by the quasidilaton models of massive gravity \cite{GannoujiHossainSamiEtal2013, AnselmiLopezNacirStarkman2015} and by 
axion models~\cite{KobayashiFerreira2018}.


\paragraph{Conclusion}
We have constrained the EoS $w$ and abundance $\omega_g$ of Dark Matter, in 8 temporal bins covering 5 decades in cosmic scale factor, 
using the CMB data from the Planck satellite, and separately including BAO and HST data. 
We found that $w$ is consistent with zero and the DM abundance is strictly positive at all cosmological epochs considered here, 
see Fig.\,\ref{varwovertime_THREE} and Fig.\,\ref{rhoovertime}, and thus the  concordance \Lcdm model remains unchallenged.
This is the first time that the level of DM ``coldness'' across cosmic time has been explicitly constrained.

\begin{acknowledgments}
The research leading to these results
has received funding from the European Research Council under the European Union's Seventh Framework Programme (FP7/2007-2013) / ERC Grant Agreement n. 617656 ``Theories
 and Models of the Dark Sector: Dark Matter, Dark Energy and Gravity''. The Primary Investigator is C. Skordis.
\end{acknowledgments}

\def\aj{AJ}
\def\aap{A\&A}
\def\apj{ApJ}
\def\aapr{A\&A Rev.}
\def\apjl{ApJ}
\def\mnras{MNRAS}
\def\araa{ARA\&A}
\def\aj{AJ}
\def\qjras{QJRAS}
\def\physrep{Phys. Rep.}
\def\nat{Nature}
\def\aaps{A\&A Supp.}
\def\apss{Ap\&SS}      
\def\apjs{ApJS}
\def\prd{Phys. Rev. D}
\def\jcap{JCAP}
\def\nar{New Astron. Rev}

\bibliographystyle{apsrev}
\bibliography{varywletter.bib}

\end{document}